\documentclass[aip,graphicx]{revtex4-1}
\usepackage{graphicx}
\usepackage{caption,mathtools,siunitx,wasysym,ulem,cancel}
\usepackage[dvipsnames]{xcolor}
\usepackage{adjustbox}
\usepackage{hhline}

\draft 
\begin{document}
\preprint{AIP/123-QED}
\captionsetup[figure]{labelfont={bf},labelformat={default},labelsep=period,name={FIG.}}
\title[Appl. Phys. Let.]{``Meta-atomless" architecture based on an irregular continuous fabric of coupling-tuned identical nanopillars enables highly efficient and achromatic metasurfaces}
\author{H. Bilge Yağcı}
\affiliation{ 
Department of Electrical and Electronics and UNAM-Institute of Materials Science and Nanotechnology, Bilkent University,TR-06800 Ankara, Turkey}
\author{Hilmi Volkan Demir}
    \email{volkan@stanfordalumni.org}
\affiliation{ 
Department of Electrical and Electronics and UNAM-Institute of Materials Science and Nanotechnology, Bilkent University,TR-06800 Ankara, Turkey}
\affiliation{Department of Physics, Bilkent University, TR-06800 Ankara, Turkey}
\affiliation{LUMINOUS! Centre of Excellence for Semiconductor Lighting and Displays, School of Electrical and Electronic Engineering, School of Physical and Mathematical Sciences, School of Materials Science and Engineering, Nanyang Technological University, 50 Nanyang Avenue, 639798 Singapore}
\date{\today}
\begin{abstract}
Metasurfaces are subwavelength-thick constructs, consisting of discrete meta-atoms, providing discretized levels of phase accumulation that collectively approximate a designed optical functionality. The meta-atoms utilizing Pancharatnam-Berry phase with polarization-converting structures produced encouraging implementations of optical components including metalenses. However, to date, a pending and fundamental problem of this approach has been the low device efficiency that such resulting metasurface components suffer, an unwanted side effect of large lattice constants that are used for preventing inter-coupling of their meta-atoms. Although the use of near-field coupling for tuning electromagnetic resonances found its use in constructing efficient narrow-band designs, such structures fell short of providing high efficiency over a broad spectrum. Here, we propose  and show that tightly packed fabric of identical dielectric nanopillar waveguides with continuously-tuned inter-coupling distances make excellent and complete achromatic metasurface elements. This architecture enables the scatterers to interact with the incoming wave extremely efficiently. As a proof-of-concept demonstration, we showed an achromatic cylindrical metalens, constructed from strongly coupled dielectric nanopillars of a single geometry as continuously-set phase elements in a “meta-atomless” fashion, working in the entirety of \SI{400}{}-\SI{700}{\nano\meter} band. This metalens achieves over 85 percent focusing efficiency across this whole spectral range. To combat polarization sensitivity, we used hexagonally stacked nanopillars to build up a polarization-independent scatterer library. Finally, a circular metalens with polarization-independent operation and achromatic focusing was obtained. This is a paradigm shift in making an achromatic metasurface architecture by wovening identical nanopillars coupled into an irregular lattice laterally constructed via carefully-tuned near-field coupling.
\end{abstract}
\maketitle
\noindent
Metasurfaces are subwavelength-thick layers of engineered structures, formed to shape wavefronts. Originally, the concept has its roots in phased array antennas, in which the signal fed to an element of the array has location-dependent phase shift with respect to the reference signal. Early examples of metasurfaces started with LC oscillators\cite{LCOscillator} and as operation frequencies grew larger, plasmonic\cite{PlasmonicMicrowave}$^{,}$\cite{PlasmonicGradientMetasurface}$^{,}$\cite{PlasmonicBabinet}$^{,}$\cite{PlasmonicTelecom} and all-dielectric designs\cite{DielectricTelecom}$^{,}$\cite{Khorasaninejad1190}$^{,}$\cite{KhorasaninejadHologram} came into being. Holograms\cite{KhorasaninejadHologram}$^{,}$\cite{ShalaevPurdueHologram}$^{,}$\cite{3DHologram}, orbital angular momentum (OAM) manipulating devices\cite{VortexBeam}$^{,}$\cite{DevlinOAM}$^{,}$\cite{FaraonOAM}, lenses\cite{Khorasaninejad1190}$^{,}$\cite{ArbabiCollimatingLens}$^{,}$\cite{Chen2018Circular}$^{,}$\cite{WangAchromaticLens} and various other functional devices\cite{ArbabiRetroref}$^{,}$\cite{AluEdgeDetection}$^{,}$\cite{CapassoSpectro} have been implemented with metasurfaces. Constituting elements, dubbed as meta-cells (or meta-atoms), function as signal processing units in space via their position-dependent phase response and vary in geometry and material with design spectra and functionality. Pancharatnam-Berry (PB) phase\cite{BerryPhase} and/or resonances in the meta-cells can be utilized to obtain the phase coverage required. Resonances in meta-cells were used in the very first metasurface designs as high-Q scattering mechanisms\cite{OAkinDielectric}. Due to the nature of resonance-tuning approach, these structures were able to offer high functional efficiency only for limited spectral band operation, and thus, narrow-band applications\cite{KuznetsovNearUnity}$^{,}$\cite{NonlocalInverse}. As the attention of photonics community shifted towards achromatic focusing, resonances became insufficient as the dominant phase contributors. Although subsequent attempts to widen operating bandwidth with multi-wavelength designs\cite{ArbabiMultiwave}$^{,}$\cite{CapassoMultiwave}$^{,}$\cite{CapassoMultiwavelength} did not suffice to justify using only resonance tuning, the reported use case of weak resonances for compensation of material dispersion\cite{Si4Dispersion} was quickly adapted into meta-atom designs and employed extensively. The PB phase, on the other hand, provided a frequency-independent phase accumulation mechanism, accompanied by an undesired polarization dependency. While it was possible to exploit polarization dependency in certain applications\cite{ARMetasurface}$^{,}$\cite{CapassoOrthPol}, achromatic operation of metalenses with the PB phase was shadowed by low polarization conversion efficiency\cite{ChenAnisotropic}, which was shown to be limited by a theoretical maximum\cite{PlasmonicMicrowave}. Efficient achromatic focusing with metalenses has remained a challenge as a result to date.\\ Due to the impressive power and flexibility of metasurface concept as an optical signal processing tool, there exists a primordial design constraint to satisfy uncoupled operation. Uncoupled operation eases the design process, but suffers two major flaws: First, the required phase function is sampled on the design to select the most suitable elements at each unit cell’s center. Commonly, the lattice constant $\Lambda$  is chosen as the inverse of Nyquist rate to faithfully construct the desired wavefront, a bare minimum when the meta-atom response perfectly matches the required functionality. Second, in analysis and optimization of the utilized scattering mechanism, inter-cell coupling is often modeled as a parasitic effect, its suppression requiring restrictions in the design space. While the idea of uncoupled operation is important for the design of spatially independent scatterers, achromatic lensing does not inherently benefit from it: the amplitude response should ideally be unity for all scatterers and the phase function required for achromatic focusing should be monotonic in space and frequency; therefore, it should be possible to design an efficient metalens with such constraints relaxed.\\ To retain achromatic operation in a wide spectral range, the dispersion between the responses of the metasurface library elements should be kept constant. In a previous study conducted in our group, dielectric pillars were shown to be suitable achromatic phase elements when operated outside of the particle resonances\cite{TanrioverIR}. Unfortunately, the effectiveness of this approach diminishes in the visible range, due to complications in the waveguide dispersion. Here, we hypothesize that strong near-field coupling between scatterers results in substantially enhanced interactivity with the incoming wave and achieves superior efficiency, while providing dispersion compensation via eliminating group dispersion delay and ensuring achromatic operation. We propose closely-coupled dielectric nanopillars with continuously-tuned interactivity as a paradigm shift in building a scatterer architecture of metasurfaces, which we coin as meta-atomless metasurfaces (MAMs) here. We intentionally avoid any internal or external resonances to guarantee achromaticity and rely only on the coupling between our nanopillars operating in guiding mode.  As a proof of concept, we show an efficient achromatic cylindrical MAM lens comprised of identical dielectric nanopillars utilizing edge-to-edge distance as the phase elements. This conceptual architecture enables 86\% focusing efficiency for \SI{400}{}-\SI{700}{\nano\meter} range, a record value across such a broadband spectrum. As the next step, we construct another library of nanopillars with honeycomb stacking, which provides us with the ability to build up a circular MAM lens working in the same spectral region with completely polarization-insensitive operation. Our highly efficient structures prove the legitimacy of coupling-tuning as a means to construct achromatic phase-accumulating elements.\\
\noindent
First, we systematically studied the effect of coupling to the field profiles of the guided modes of the nanopillars. With decreasing inter-pillar distance, weakly-guided modes of the uncoupled nanopillars start to interact with each other, increasing the field amplitudes in the surrounding medium. Such increase in field amplitudes can be thought as an additional index, justified by the effective index formulation for a guided mode\cite{SnyderOpticalWaveguides}. In turn, this coupling-dependent effective index change allows us to form phase elements for MAM. Additionally, as the fields are localized in the gap, the coupled operation increases the interactivity of the scatterers with the incoming wave, effectively increasing their scattering efficiency. \\Figure \ref{fig:Fig1}.(a) shows an illustration of the scatterers, with $h$ being the height of each nanopillar and $d_1$ and $d_2$ denoting the edge-to-edge distances between the nearest nanopillars in the respective directions along and perpendicular to the input polarization. 
Figures \ref{fig:Fig1}.(b-d) visualize electric field profiles along the input polarization for $\lambda=$\SI{700}{\nano\meter}, normalized to the field maxima for each case. In the uncoupled case (Figure \ref{fig:Fig1}.(b)), nanopillars hardly interact with each other. When nanopillars are coupled along the input polarization (dubbed parallel-coupled case), shown in Figure \ref{fig:Fig1}.(c), interactions between nanopillars are visible and fields are confined in a smaller region, a precursor of field enhancement or increase in effective index. Similarly, coupling nanopillars orthogonal to the input polarization (dubbed orthogonal-coupled case) in Figure \ref{fig:Fig1}.(d) creates localization, and while the mode is more confined than the uncoupled case, it is weaker than the parallel-coupled case. Such phenomenon can be explained with breaking of mode degeneracy, observed in nanopillar dimers\cite{BirefSiNanowire}.\\ 
To test our hypothesis, we next studied periodic nanopillars with fixed $r=\SI{45}{\nano\meter}$ and $d_2=\SI{30}{\nano\meter}$ while $d_1$ was used as an independent parameter. The obtained effective index map for the parameter space can be seen in Figure \ref{fig:Fig2}.(a). Using the parallel-uncoupled case as a reference, we obtained an index difference map that unfolds achromatic nature of the coupling (Figure \ref{fig:Fig2}.(b)). Although a small dispersion mismatch exists towards short-wavelength limit that stems from the unbalanced compensation of dispersion for $d_2=\SI{30}{\nano\meter}$, this represents the best trade-off between flatness and magnitude of index difference. These index difference results can be translated to the phase differences acquired in $\Delta{}h$ distance by \( \displaystyle \Phi = 2\pi\mathrm{n}_{eff}\frac{\Delta h}{\lambda} \). To validate the waveguiding approach, the responses of the nanopillars were also obtained via finite-difference time-domain (FDTD) computations. In Figure \ref{fig:Fig2}.(c), optical transmission through the scatterer is shown over the wavelength with varying edge-to-edge distance.  High transmission throughout the parameter space stems from the lack of cutoff for the fundamental mode in cylindrical waveguides and high coupling of the incident wave to the guided modes; small ripples in the transmission can be thought of reflection from an effective medium slab. Minimum for the transmission does not fit to this explanation and is mostly a consequence of dipole resonances. The phase difference variation over $\lambda$ with respect to $d_1$ (Figure \ref{fig:Fig2}.(d)) exhibits a great similarity to Figure S1, confirming the validity of our approach.\\
To verify the applicability of rectangular-packed structures, here we implemented a cylindrical MAM lens with $f=\SI{32}{\micro\meter}$ and NA = 0.154, following the formulations of the phase requirements of a cylindrical lens presented in the Supporting Information. 
Utilizing Equation S4, we acquired an $r$-dependent edge-to-edge distance map, which is used for placing identical nanopillars on the lens plane (Figure \ref{fig:lensing}.(a)). At each axial position, the required phase $\Phi_{req}$ is sampled by the actual imposed phase $\Phi_{imp}$ with consistency over the design spectra (Figure \ref{fig:lensing}.(b)). As shown in Figure \ref{fig:lensing}.(c), the proposed structure focuses efficiently across the entire bandwidth, with the focusing efficiencies over 85\% and the absolute efficiency levels over 80\%. False-colored images of intensities along the optical axis of the lens are displayed in Figure \ref{fig:lensing}.(d), normalized with respect to the intensity maxima at each wavelength. 
While our architecture is diffraction-limited for the short-wavelength limit, its ability to resolve finer details suffers as the wavelength increases. The extended depth-of-focus for all wavelengths allows one to optimize for a more uniform resolution at all wavelengths without a significant loss in the focal spot intensity.\\
\noindent
Tuning the coupling in only one direction, as we did in our cylindrical MAM lens, imposes a polarization dependency to the design (see the Supporting Information for further discussion). To combat this, we can also stack the nanopillars in a honeycomb lattice. While a honeycomb lattice of nanopillars is not 4-fold symmetric specifically, its response approximates a circularly symmetric scatterer as the length scales shrink, allowing to design polarization-insensitive scatterers without losing performance. As shown in Figures \ref{fig:hexphase}.(a-f), a honeycomb lattice with $r=\SI{45}{\nano\meter}$ has similar qualities to the rectangular-coupled nanopillars in Figure \ref{fig:Fig2}: it possesses a similarly flat dispersion curve apparent from the almost $\lambda$-independent nature of the effective index difference map shown in Figure \ref{fig:hexphase}.(b). Effective index differences between orthogonal excitations are negligible with $<1\%$ difference for most of the spectrum as seen in Figure \ref{fig:hexphase}.(c). Figures \ref{fig:hexphase}.(d-e) show a comparison between the waveguide phase (Fig. \ref{fig:hexphase}.(d)), calculated from the effective index map, and the phase response acquired via FDTD computations (Fig. \ref{fig:hexphase}.(e)). The striking similarity of two approaches in terms of the shape and the magnitude of the phase response supports the validity of our approximation. Honeycomb lattice is also highly transmittive, as seen in Figure \ref{fig:hexphase}.(e). Subsequently, we designed a circular, achromatic and polarization-insensitive MAM lens with NA = 0.26 and $f=\SI{11}{\micro\meter}$. A relatively small lens diameter was chosen to impose a fine mesh while respecting computational limitations. To minimize the placement errors originating from the tiling of our irregular lattice, we utilized an optimization algorithm  (see the Supporting Information for details). The resultant architecture and its characteristics are shown in Figure \ref{fig:hexlens}.\\
As seen in Figure \ref{fig:hexlens}.(a), phase requirements for this metalens constituted of such a continuously tuned lattice of inter-coupled nanopillars as described above are met with little to no mismatch, with $\lambda=\SI{550}{\nano\meter}$ presenting the best match. Focusing characteristics under orthogonal input polarizations (Figure \ref{fig:hexlens}.(b)) exhibit a small contrast between X- and Y-polarized cases. Instead of presenting absolute efficiency, total transmission from the lens is portrayed to avoid congestion in the figure. The mismatch might be the result of a slight disturbance in the 4-fold symmetry of the nanopillar placement on lens plane. Relatively low focusing efficiency compared to cylindrical lens is caused by relatively strong side-lobes, visible in the focal spot profiles. $\lambda$-specific beam patterns and Strehl ratios are displayed in Figures \ref{fig:hexlens}.(c-d) with false coloring. As seen in Figure \ref{fig:hexlens}.(c), the focal spot of the obtained metalens is invariant to changes in wavelength, while its beam width and depth of focus have a linear dependence. This metalens also features superior resolution, seen in Figure \ref{fig:hexlens}.(d). Diffraction-limited values for Strehl ratio are reached for all wavelengths, with relatively enhanced performance in the long-wavelength limit. This is attributed to a smaller phase mismatch at longer wavelengths.\\
In Table \ref{tab:comparison}, we compare our achromatic MAM lenses with the recent works reported in the field\cite{Chen2018Circular}$^,$\cite{WangAchromaticLens}$^,$\cite{ChenAnisotropic}$^,$\cite{MultipleShapes}$^,$\cite{Fishnet}. The proposed cylindrical MAM lens outperforms all the other designs in efficiency, whereas the circular MAM lens is highly diffraction-limited, achromatic and has polarization-insensitive operation. Here, the high absolute efficiencies of our devices prove the effectiveness of our MAM paradigm in wavefront manipulation.\\
In summary, we have proposed and demonstrated that a continuously tuned fabric of inter-scatterer coupling can be utilized as an achromatic phase accumulation mechanism for geometries operating in guiding mode. This paradigm allows us to devise metasurfaces that interact with the incoming wave much more effectively than metalenses of similar functionalities previously reported in the literature, all of which rely on uncoupled meta-cells. Here, based on our MAM concept, we were able to show polarization-insensitive and efficient operation and achieve fully achromatic focusing in the visible range. While the fabrication process for the designs are relatively demanding, this approach can be conveniently applied, for example, to the infrared region, where the length scales in question are easily reachable with electron-beam lithography. The same challenges may be resolved by utilizing a dielectric of higher refractive index as the pillar material and relaxing the high aspect ratio requirements as the same phase is acquired in a smaller distance. The findings here altogether indicate that continuously tuning the coupling of identical nanopillars in a two-dimensional lattice provides a highly effective approach to make efficient metasurfaces. 
\\
See the Supporting Information for simulation and implementation details of MAM lenses, as well as the definitions of figure-of-merits used in the text.
\begin{acknowledgments}
The authors gratefully acknowledge the financial support in part from Singapore National Research Foundation under the programs of NRF-NRFI2016-08, NRF-CRP14-2014-03, and the Science and Engineering Research Council, Agency for Science, Technology, and Research (A*STAR) of Singapore and in part from TUBITAK 115F297, 117E713, and 119N343. H.V.D. gratefully acknowledges support from TÜBA. Also, the authors thank Mr. İbrahim Tanrıöver for his assistance in the early phase of this work, in which he developed another implementation for the concept and ideas of this work proposed by H.V.D. H.B.Y. also thanks Mr. Alper Ahmetoğlu and Mr. Tevfik Bülent Kanmaz for fruitful discussions in stacking optimization.
\end{acknowledgments}
\section*{DATA AVAILABILITY}
\noindent
The data that support the findings of this study are available from the corresponding author upon reasonable request.
\bibliography{24Jan21.bib}
\newpage
\begin{figure}
    \centering
    \includegraphics[scale=0.5]{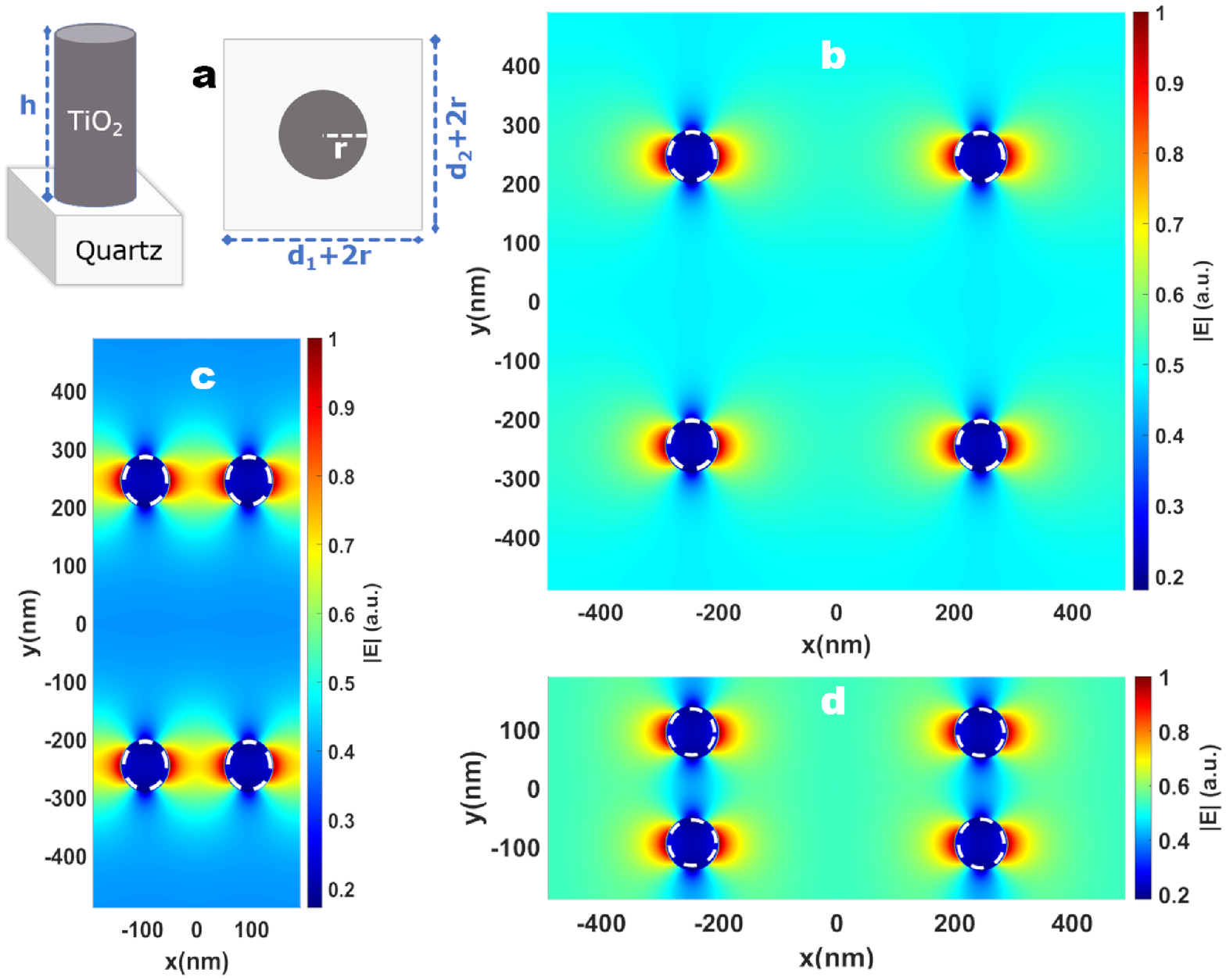}
    \caption{\label{fig:Fig1} (a) An exemplary scatterer, defined with geometrical parameters $r$, $h$ and lattice parameters $d_1$,$d_2$. (b-d) Horizontal slice of electric field components along the input polarization for nanopillars ($\{r,h\}=\{\SI{45}{\nano\meter},\SI{600}{\nano\meter}\}$) in specified lattices. Nanopillars are superimposed on the fields with white lines. To visualize field coupling better, simulated periodic cell region is constructed from four closely located nanopillars.  (b) Uncoupled case ($\{d_1,d_2\}=\{\SI{400}{\nano\meter},\SI{400}{\nano\meter}\}$). (c) Coupled along the input polarization (parallel-coupled) ($\{d_1,d_2\}=\{\SI{100}{\nano\meter},\SI{400}{\nano\meter}\}$). (d) Coupled orthogonal to the input polarization (orthogonal-coupled) ($\{d_1,d_2\}=\{\SI{400}{\nano\meter},\SI{100}{\nano\meter}\}$). }
    \label{fig:Fig1}
\end{figure}
\begin{figure}
    \centering
    \includegraphics[scale=0.5]{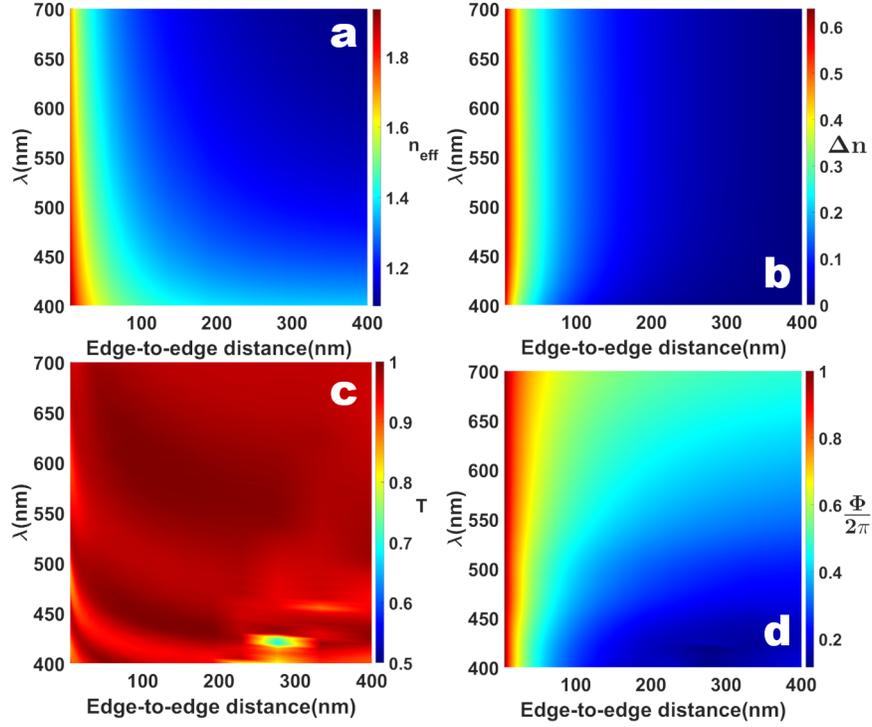}
    \caption{\label{fig:Fig2} Numerical simulation results for a lattice of $\mathrm{TiO}_2$ nanopillars. Stacking is along the input polarization. Edge-to-edge distance in the orthogonal direction is \SI{30}{\nano\meter}. (a-b) Simulation results for an array of infinite cylinders suspended in vacuum, done in MODE Waveguide Simulator. (a) $\mathrm{n}_{eff}$ versus $\lambda$ and $d_1$ for the guided mode, polarized along the input polarization. (b) Effective index difference map. For all $\lambda$, parallel-uncoupled case is taken as a reference. Index difference is constant for changing wavelength for most of the spectrum, implying achromaticity. (c-d) FDTD results for \SI{600}{\nano\meter}-high $\mathrm{TiO}_2$ nanopillars, residing on quartz substrate, with the same lattice setup in (a-b). (c) Transmission map of the nanopillars. (d) Phase response of the nanopillars.}
    \label{fig:Fig2}
\end{figure}
\begin{figure}
    \centering
    \includegraphics[scale=0.5]{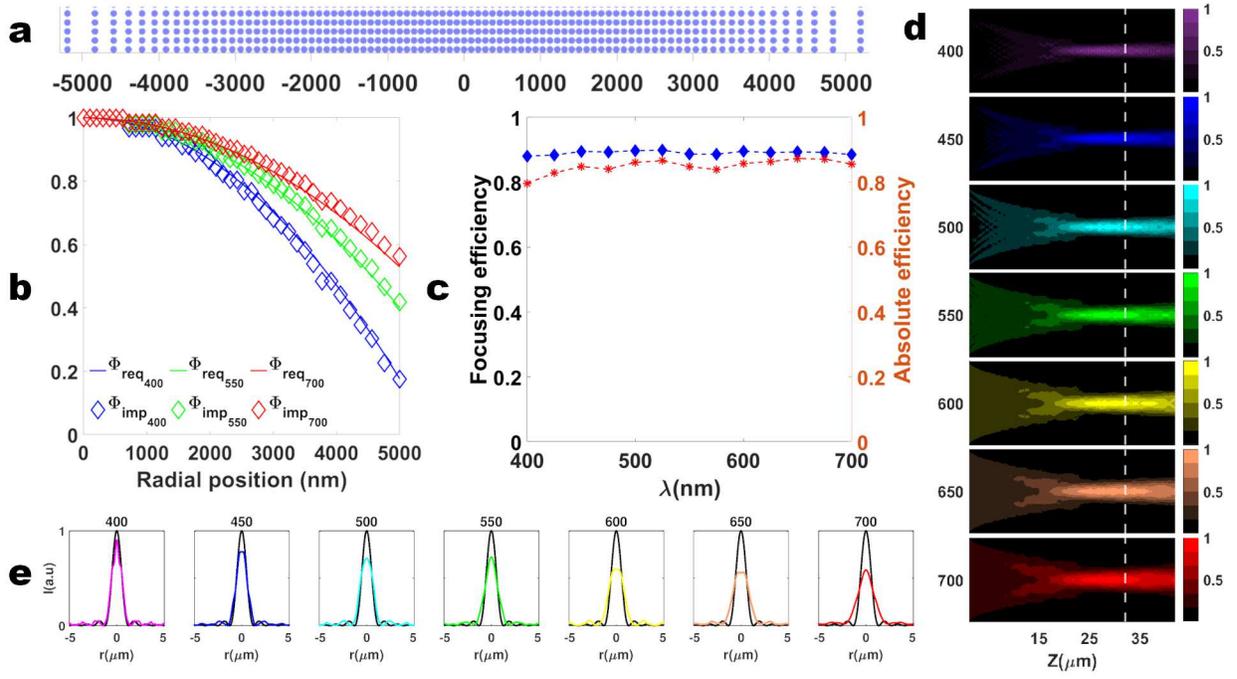}
    \caption{ \label{fig:lensing} Design and characterization of the cylindrical MAM lens. (a) Placement of nanopillars on the design plane. Variation in $d_1$ is used for constructing phase elements, while $d_2$ is constant with $\SI{30}{\nano\meter}$, $r=\SI{45}{\nano\meter}$. (b) Required phase response for the design $\Phi_{req_{\lambda}}$ and realized phase response $\Phi_{imp_{\lambda}}$ for $\lambda=\{\SI{400}{\nano\meter},\SI{550}{\nano\meter},\SI{700}{\nano\meter}\}$. (c) Relative and absolute focusing efficiency of the design. Both efficiencies are above 80\% over the spectral band, showing the capability of the proposed structure as a highly-efficient metalens. (d) False-colored image of normalized intensities along the optical axis. White dashed line shows the design focus at $z=\SI{32}{\micro\meter}$. (e) Wavelength-specific Strehl ratios, normalized to slit diffraction pattern. Dashed line marks a Strehl ratio of 0.8.
    } 
    \label{fig:lensing}
\end{figure}
\begin{figure}
    \centering
    \includegraphics[scale=0.5]{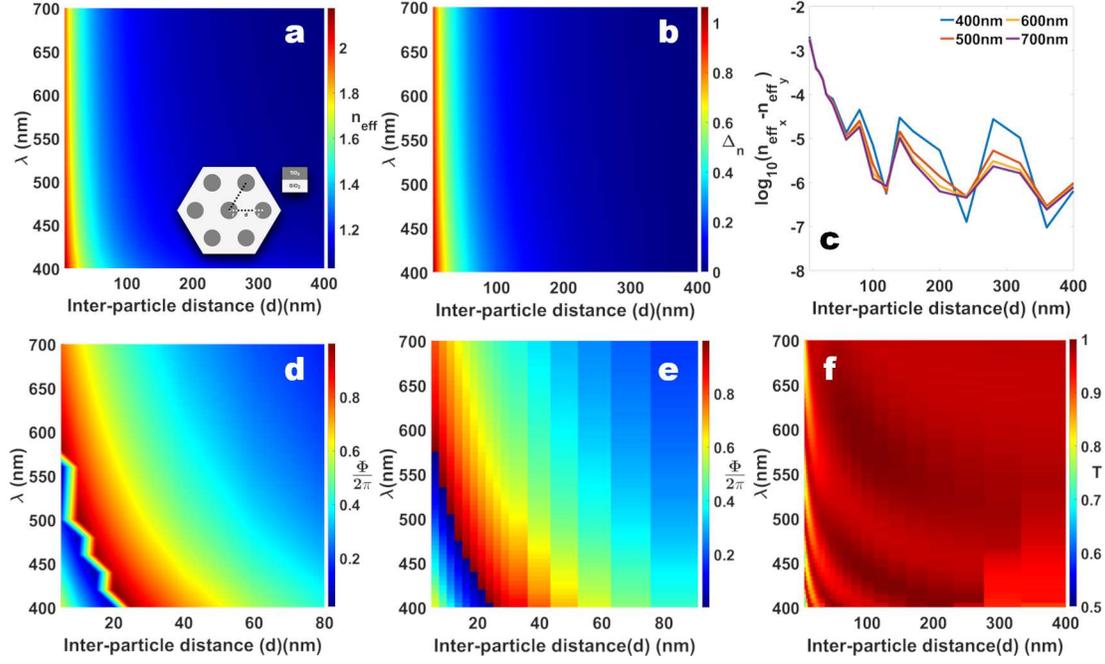}
    \caption{ \label{fig:hexphase} Phase response and transmission pattern of nanopillars stacked in honeycomb lattice. (a-d) MODE simulations for nanopillars stacked in honeycomb lattice, with $r=\SI{45}{\nano\meter}$ and inter-particle distance $d$ chosen as the independent parameter.  (a) Effective index map for honeycomb stacking. Excitation polarization is chosen parallel to the horizontal direction. Inset: illustration of nanopillars in a honeycomb lattice. (b) Effective index difference map for the geometry used in part (a). Higher effective index difference compared to the rectangular stacking of nanopillars follows from a relatively high fill factor and coupling strength. (c) Effective index differences between the horizontally-polarized and the vertically-polarized excitation. For all design spectra, index difference between these polarizations is less than 5\permil. (d) Calculated phase response map for the lattice in (a), obtained via \( \displaystyle \Phi = 2\pi\mathrm{n}_{eff}\frac{\Delta h}{\lambda} \) with $\Delta h = \SI{600}{\nano\meter}$, normalized to the maximally uncoupled case ($d=\SI{400}{\nano\meter}$). (e-f) FDTD simulation results for $\SI{600}{\nano\meter}$-high nanopillars standing on a quartz substrate. Input excitation is aligned with the horizontal direction. (e) Phase response map of the honeycomb lattice, normalized to the maximally uncoupled case. (f) Transmission map of the honeycomb lattice.
    } 
    \label{fig:hexphase}
\end{figure}

\begin{figure}
    \centering
    \includegraphics[scale=0.5]{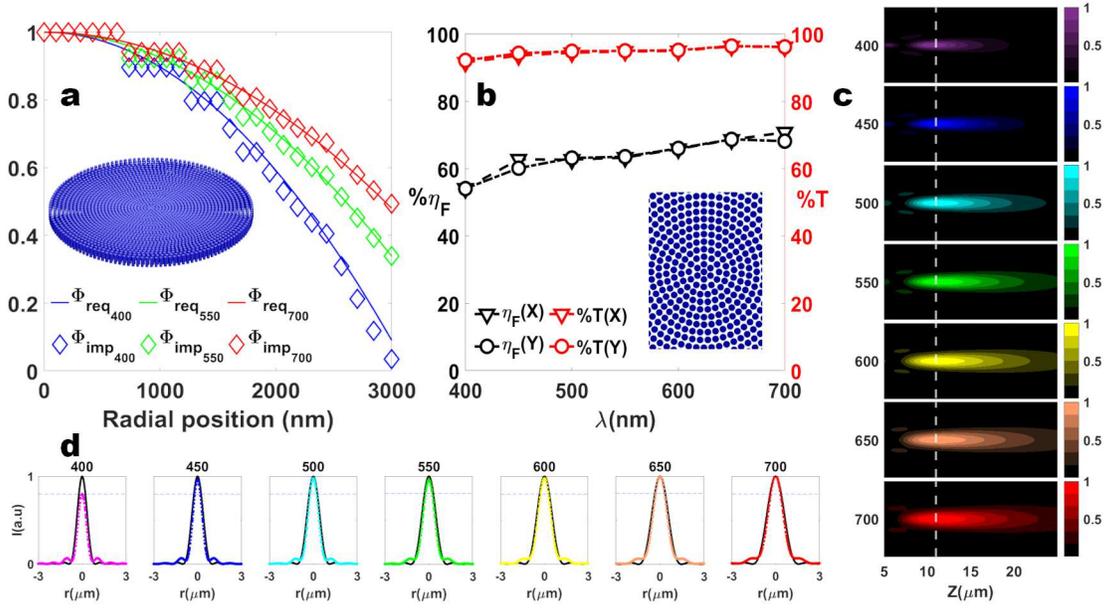}
    \caption{ \label{fig:hexlens} Proof of concept demonstration of a circular, achromatic and polarization-insensitive lens, with NA = 0.26 and $f=\SI{11}{\micro\meter}$, implemented based on MAM paradigm. (a) Phase requirements of the lens at the band edges and the mid-band, shown with $\Phi_{req_{\lambda}}$, plotted alongside the realized phase response at the same wavelengths, $\Phi_{imp_{\lambda}}$. Inset: Perspective view of a circular lens. (b) Focusing efficiencies $\eta_F$ and total transmissions for X and Y-polarized excitations. Total transmissions from the monitors can be translated to the absolute efficiencies by applying $\eta_{abs_{x,y}} = \eta_{F_{x,y}}*T_{x,y}$. Inset: Close-up image of the metalens in the central region. (c) False-colored image of the normalized intensities along the optical axis. White dashed line annotates the focal plane at $f=\SI{11}{\micro\meter}$. (d) Wavelength-specific Strehl ratios, calculated according to the diffraction pattern from a circular aperture. Dashed line marks the diffraction-limited value of Strehl ratio (0.8). Relatively large deviations from high Strehl ratios in short-wavelength limit may be caused by the larger $\Phi_{req_{400}}$-$\Phi_{imp_{400}}$ mismatch, shown in (a). 
    }
    \label{fig:hexlens}
\end{figure}
\begin{table}
    
    \caption{ Performance metrics of previously reported achromatic metalenses and their comparison to our MAM lenses. }
        
\begin{adjustbox}{width={\textwidth},totalheight={\textheight},keepaspectratio}

    \begin{tabular}{ccccccc}
    \hhline{=======}
         &$\frac{\text{FWHM spot size(}f_{center}\text{)}}{\text{Diffraction limit(}f_{center}\text{)}}$& Strehl ratio & $\eta_{abs}(f_{center})$ &Polarization&$\frac{\Delta f}{f_{center}}$& N.A.   
         \\ Reference 15\footnotemark[1] & $\sim\SI{1.35}{\micro\meter}/\SI{1.35}{\micro\meter}$  & $\sim$90\% & $\sim$20\% & Circular & 190 THz/525 THz & 0.2 
         \\ Reference 16\footnotemark[1] & $\sim\SI{2.75}{\micro\meter}/\SI{2.5}{\micro\meter}$ &$\times$ & $\sim$50\% & Circular & 195 THz/565 THz & 0.106 
         \\ Reference 30\footnotemark[1] & $\sim\SI{1.5}{\micro\meter}/\SI{1.46}{\micro\meter}$ & $\sim$80\% & 34\% & Insensitive & 225 THz/515 THz & 0.2
         \\ Reference 34\footnotemark[1] & $\sim\SI{800}{\nano\meter}/\SI{760}{\nano\meter}$ & N.A. too high & $\times$ & Insensitive & 35 THz/230 THz & 0.88
         \\ Reference 35\footnotemark[1] & $\sim\SI{4.28}{\micro\meter}/\SI{4.28}{\micro\meter}$ & $\sim$85\% & 65\% & Insensitive & 218 THz/325 THz & 0.12
         \\ Cylindrical MAM lens\footnotemark[2] & $\SI{1.87}{\micro\meter}/\SI{1.78}{\micro\meter}$ & 73\% & 86\% & Linear & 320 THz/545 THz & 0.154
         \\ Circular MAM lens\footnotemark[2] & $\SI{1.06}{\micro\meter}/\SI{1.03}{\micro\meter}$ & 96\% & 62\%  & Insensitive & 320 THz/545 THz & 0.26
         \\

      \hhline{=======}       
    \end{tabular}

    \footnotetext[1]{Measurement}
    \footnotetext[2]{Simulation}
    \footnotetext[24]{Not available}

\end{adjustbox}
\label{tab:comparison}

\end{table}

\end{document}